# "Draw me a curator" Examining the visual stereotyping of a cultural services profession by generative AI

Dirk HR Spennemann [1,2]*

[1]  School of Agricultural, Environmental and Veterinary Sciences, Charles Sturt University; Albury NSW 2640, Australia

[2]  Libraries Research Group; Charles Sturt University; Wagga Wagga NSW 2678, Australia

*   Correspondence: dspennemann@csu.edu.au

**Abstract**

Based on 230 visualisations, this paper examines the depiction of museum curators by the popular generative Artificial Intelligence (AI) model, ChatGPT4o. While the AI-generated representations do not reiterate popular stereotypes of curators as nerdy, conservative in dress and stuck in time rummaging through collections, they contrast sharply with real-world demographics. AI-generated imagery extremely underrepresents women (3.5% vs 49%–72% in reality) and disregards ethnic communities other than Caucasian (0% vs 18%–36%). It only overrepresents young curators (79% vs approx. 27%) but also renders curators to resemble yuppie professionals or people featuring in fashion advertising. Stereotypical attributes are prevalent, with curators widely depicted as wearing beards and holding clipboards or digital tablets. The findings highlight biases in the generative AI image creation dataset, which is poised to shape an inaccurate portrayal of museum professionals if the images were to be taken uncritically at 'face value'.

## 1. Introduction

There will be no exhibitions, no public education and no preservation of humanity's cultural heritage without the work of curators who by and large work behind the scenes. How does the general public perceive them and their roles? While there is a considerable body of work on the nature and dangers of stereotyping narratives, objects and their creators in museum exhibitions [1-3], there only a surprisingly limited and incoherent body of work that considers the cultural stereotypes of museum curators themselves. A systematic review found only a handful of references, with only commenting specifically on the topic, one of which is dated [4] and one of which is a blog post [5]. In that curators differ from other cultural industry professions such as archivists [6, 7] and librarians [8-10].

The pervading stereotypes of a curator, largely driven by ignorance of their actual work by the broader public, are that of 'icy, turtlenecked snobs or puttering old men toiling away in dusty tombs' [11]. They most commonly are seen as stuck in time, boring, dusty and old [12] or dusty and dry [13] persons. They are 'nerdy, conservative in dress and wear glasses" [4], commonly being a bearded, white, male … shuffling through dusty boxes curating specimens [14] in the basement. Other stereotypes of curators reflect the perceptions of the spaces they are working in, such as "dusty, gloomy, and nostalgic places frozen in time [15] An analysis of representations of curators in pop-culture stereotyped them either as an art curator with desirable attributes of glamour, power, and avantgarde, whereas "museum curator characters are all stuffiness, dogma, and tweed" [5]. Looking at popular representations in literature and film, 62% of curators are portrayed as women, with 89% being white/Caucasian [5].

Stereotypes are oversimplified and often inaccurate generalizations about groups, ignoring individual differences. Professions and their attire are frequently stereotyped (e.g., "men are builders," "scientists wear lab coats"). Gender and ethnic stereotypes in careers can harm self-esteem, limit opportunities, and lead to discrimination. As a result, those who do not fit these moulds may avoid certain professions, unintentionally reinforcing these biases [16]. Stereotypes, which form early, often in childhood and which become difficult to change once firmly entrenched, often are reinforced through visual imagery in advertisements, movies, and social media [17, 18].

The recent development of generative AI text to image models such as DALL-E , Stable Diffusion or Midjourney has enabled the rapid creation of copyright-free, realistic images for various uses, reducing reliance on stock photos. Generative AI models cannot be entirely free from bias due to factors like model design, training data quality, and developer influence during design and training. Bias may stem from ideological leanings, the choice of language when training, or use of sources which may contain outdated or racially biased information [19-21]. Studies have shown that AI-generated responses and images frequently depict professions in gender-stereotypical ways. A cross-professional study found that 60% of AI-generated professions aligned with societal stereotypes (curators were not assessed) [22]. Bias is also evident in text-to-image AI programs like DALL-E, where prompts often lead to underrepresentation or misrepresentation of gender and ethnicity [23, 24]. When gender is unspecified, generative AI tends to depict men in authoritative roles more often than women [8, 25]. Similarly, AI-generated images predominantly feature Caucasian individuals, while African-American characters are often shown in service roles [24]. These biases extend to physical appearance, favouring young individuals and adhering to stereotypical beauty norms. Notably, images of pregnant women or people with disabilities are largely absent in single-shot AI-generated prompts, highlighting the lack of diverse representation in AI-generated visuals [25].

This paper examines experimentally how a popular generative AI model, ChatGPT4o, interprets and visualises curators in four museum settings (art, fashion, natural history, social history).

## 2. Methodology

The prompts provided to ChatGPT4o were deliberately formulated in an unconstrained fashion to avoid responses that were biased towards the user's perceptions. ChatGPT4o itself generated the complex textual prompt for DALL-E to render the image. No human manipulation in the image conceptualisation and implementation process occurred. The following prompt was used: "*Think about [insert museum type] museums and the curators working in these. Provide me with a visualisation that shows a typical curator against the background of the interior of the museum.*" Each resulting image was saved to disk, the AI-generated prompt retrieved from the image panel and, together with the image, copied into a data file. After saving, the chat was deleted to ensure a clean and unbiased generation of a new image without legacy information available to ChatGPT4o. Fifty images each were generated for each of the four museum types 'art museum' (AM), 'fashion museum' (FA), 'natural history museum' (NH) and 'social history museum' (SH). In addition, ten images each were generated for the museum types 'maritime museum' (MA), 'science museum' (SC) and 'technology museum' (TE).

Scored were visual cues to gender (male/female) and age, with the latter scored as young, middle aged and old. The criteria boundaries for age were somewhat fluid, drawing solely on facial features and hair colouring. Additionally scored were the presence of a beard for males and the hair style for women (bob, open long hair, bun, ponytail) as well as the presence of spectacles, a book, tablet or clipboard. The museum setting in the background was scored in terms of nature of exhibits, the positioning of windows and the nature of showcases. Summary data and frequencies were established using MS Excel, while the statistical comparisons were established with MEDCALC's comparison of proportions calculator [26].

The images were generated on 9–11 March 2025 [27]. The images as well as the prompts that ChatGPT4o used to generate the images have been archived in a supplementary data file according to an established protocol [27] at the author's institution and can be accessed via doi: 10.26189/951bb16a-3dda-4a84-bd3d-994203d28c7e.[28].

## 3. Results

Of the total 230 images of curators generated for this study, the overwhelming majority (96.5%) depicted men (Table 1). All the eight women curators rendered by DALL-E are young, professionally dressed in an ensemble. None are wearing glasses or name tags. Their hair is worn long and open (4), tied in a bun (2), tied back, or cut short. Given the small number of women, the remainder of the discussion will address the portrayal of male curators.

The majority of male curators are depicted as young men (Table 1), ranging from 100% among curators in fashion museums to 74% of curators in social history museums. The museums cluster onto three groups: fashion, natural history and science; art and social history; and maritime and technological. Art and social history) have a significantly greater proportion of middle-aged staff than fashion or natural history (AM vs. NH $\chi^2$=4.921, df=1, $p$=0.0265; SH vs. NH $\chi^2$= 5.683, df=1, $p$=0.0171), while curators in maritime and technological museums are significantly even older than the curators in social history (SH vs. MA $\chi^2$=10.503, df=1, $p$=0.0012; SH vs. TE $\chi^2$= 7.139, df=1, $p$=0.0075).

The generative text-to-image combination of ChatGPT and DALL-E generated several persistent curator attributes. For example, more likely than not, curators in natural history, maritime and science museums are wearing name tags or lanyards in the images rendered by ChatGPT / DALL-E, while they are eschewed altogether by curators working in art and fashion museums (Table 1). If the depictions by ChatGPT / DALL-E were to be believed, it is *de rigeur* for museum curators to walk through their exhibits armed with a clipboard (and notebooks) or digital tablets. Clipboards are very 'popular' among curators in art and fashion museums (96–98%), while tablets dominate in technological museums (90%) (Table 1). Given that beards are a stereotypical attribute of a male scientist [14, 29, 30] it is not surprising that three quarters of more curators in art, natural history and maritime museums are depicted in that way. The exception are curators in fashion museums, where facial hair is depicted in only 20.9% of instances. (Table 1). Glasses (spectacles), another common stereotypical attribute associated with learnedness and scientists [31-33], are not depicted with any coherent logic. They are particularly common in the portrayals of curators of technical museums as well as art museums (71.4%–80%) and less common among curators working in science and fashion museums (30%–34.9%) (Table 1).

*Table 1. Characteristics of male curators depicted by ChatGPT4o*
*Museum type: AM–art ; FA–fashion ; MA–maritime ; NH–natural history ; SC –science ; SH–social history ; TE.–technology*

|  | AM | NH | SH | FA | MA | SC | TE |
|---|---|---|---|---|---|---|---|
| **Gender** | | | | | | | |
| Women | 2.0 | — | — | 14.0 | — | — | — |
| Men | 98.0 | 100.0 | 100.0 | 86.0 | 100.0 | 100.0 | 100.0 |
| n | 50 | 50 | 50 | 50 | 10 | 10 | 10 |
| **Age class** | | | | | | | |
| young | 75.5 | 92.0 | 74.0 | 100.0 | 20.0 | 80.0 | 30.0 |
| middle age | 24.5 | 8.0 | 26.0 | — | 70.0 | 20.0 | 70.0 |
| old | — | — | — | — | 10.0 | — | — |
| n | 49 | 50 | 50 | 43 | 10 | 10 | 10 |
| **Attributes** | | | | | | | |
| glasses | 71.4 | 42.0 | 40.0 | 34.9 | 50.0 | 30.0 | 80.0 |
| beard | 75.5 | 76.0 | 68.0 | 20.9 | 90.0 | 60.0 | 50.0 |
| name tag | — | 66.0 | 24.0 | — | 70.0 | 80.0 | 40.0 |
| n | 47 | 50 | 50 | 43 | 10 | 10 | 10 |
| **Objects** | | | | | | | |
| book | — | — | — | 4.0 | — | — | — |
| clipboard | 96.0 | 62.0 | 70.0 | 88.0 | 40.0 | 68.0 | — |
| clipboard & camera | — | — | — | — | — | 2.0 | — |
| notebook | — | 4.0 | — | 8.0 | — | — | — |
| object | — | 2.0 | 20.0 | — | — | 2.0 | — |
| tablet | 2.0 | 6.0 | — | — | 10.0 | 8.0 | 90.0 |
| none | 2.0 | 26.0 | 10.0 | — | 50.0 | 20.0 | 10.0 |
| n | 50 | 50 | 10 | 50 | 10 | 50 | 10 |

Curators in art museums are significantly more likely to wear informal clothing, such as a blazer and pants (Figure 1b) than formal wear, such as suits (with or without vests) (Figure 1c) ($\chi^2$= 12.83, df=1, $p$=0.0003), as are curators in social history museums ($\chi^2$= 5.702, df=1, $p$=0.0169)( Table 2). Curators working in natural history museums are significantly more likely to be depicted as wearing formal clothing than laboratory coats ($\chi^2$= 17.813, df=1, $p$<0.0001) (Figure 1a) or casual attire ($\chi^2$= 26.009, df=1, $p$<0.0001) (Table 2).

*Table 2. Attire worn by curators depicted by ChatGPT4o*
*Museum type: AM–art ; FA–fashion ; MA–maritime ; NH–natural history ; SC –science ; SH–social history ; TE.–technology*

|  | AM | FA | MA | NH | SC | SH | TE |
|---|---|---|---|---|---|---|---|
| **Formal** | | | | | | | |
| ensemble, open shirt | — | 10.0 | — | — | — | — | — |
| suit, vest, shirt & tie | 20.0 | 10.0 | 30.0 | 32.0 | — | 18.0 | 10.0 |
| suit, shirt & tie | 10.0 | 20.0 | 10.0 | 2.0 | 10.0 | 8.0 | 20.0 |
| vest, shirt & tie | — | 2.0 | — | 30.0 | 10.0 | 10.0 | — |
| shirt & tie | 2.0 | — | — | — | 20.0 | 2.0 | 20.0 |
| **Informal** | | | | | | | |
| jacket, open shirt | 42.0 | 42.0 | 30.0 | 10.0 | 60.0 | 42.0 | 40.0 |
| jacket, vest/sweater, open shirt | 26.0 | 14.0 | 10.0 | — | — | 16.0 | 10.0 |
| jacket, vest & shirt | — | 2.0 | 20.0 | — | — | 2.0 | — |
| open shirt | — | — | — | 4.0 | — | 2.0 | — |
| **Laboratory** | | | | | | | |
| laboratory coat, open shirt | — | — | — | 4.0 | — | — | — |
| laboratory coat, shirt & tie | — | — | — | 18.0 | — | — | — |
| n | 50 | 50 | 10 | 50 | 10 | 50 | 10 |

Although not the focus of the research, the wording of the prompt requested ChatGPT4o to place the curators into a museum setting which permits to examine how ChatGPT4o/DALL-E constructs these in terms of exhibitions (Table 3) and architecture (Table 4). Most of the museum buildings, except for science and technology museums (Figure 1d), are predominantly depicted as hallowed halls of the nineteenth century (Figure 1a–c). Some visualisations show visitors in the background. Their presence ranges from entirely absent in fashion museums to being in 90% of all depictions in science and technology museums (Table 4).

*Table 3. Museums: exhibition content depicted by ChatGPT4o*
*(percentages are given in brackets)*

| Museum type | dominant exhibition content |
|---|---|
| Art | classical paintings & statues (96); classical paintings (2); modern art (2) |
| Fashion | mannequins (100) |
| Maritime | boats and maritime paraphernalia (100) |
| Natural history | dinosaur skeletons (100) |
| Science | dinosaur skeleton, space (60); dinosaur skeleton, physics (20); dinosaur skeleton, space & DNA (10); robot, space (10) |
| Social History | costumes, paintings, objets d'art (62); paintings, objets d'art (32); costumes, text panels (4); objets d'art (2) |
| Technology | robot, CRT monitors, screen (50); robot, technopunk, screen (30); technopunk, CRT monitors, screen (10); technopunk, screen (10) |

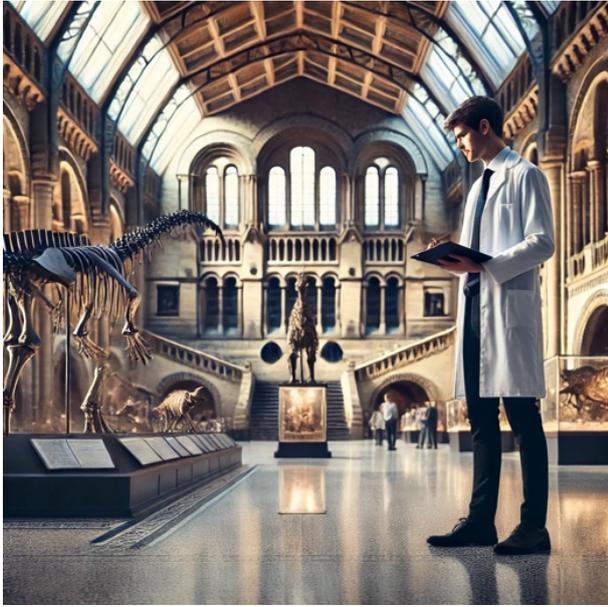 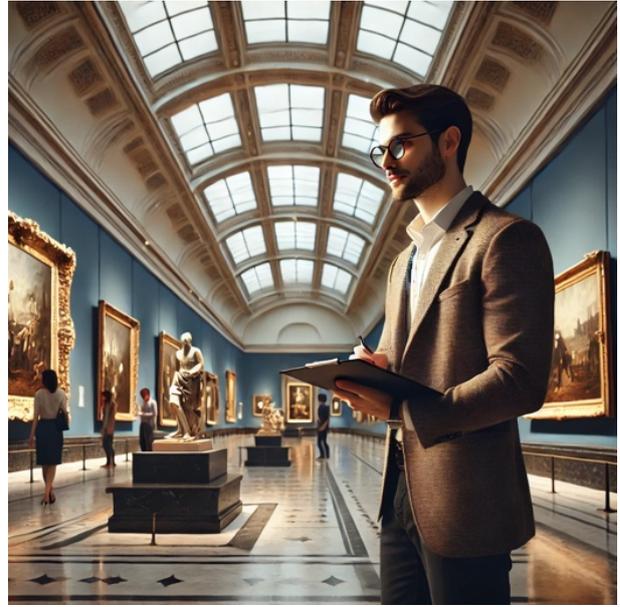

a) curator in a natural history museum (NH17)    b) curator in an art museum (AM4)

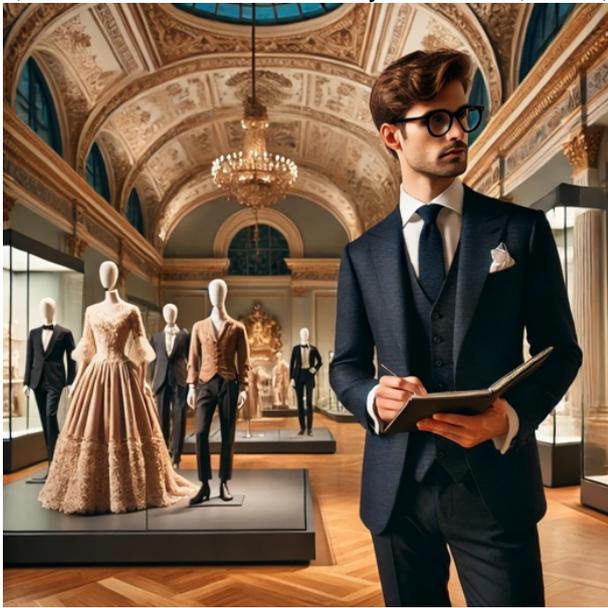 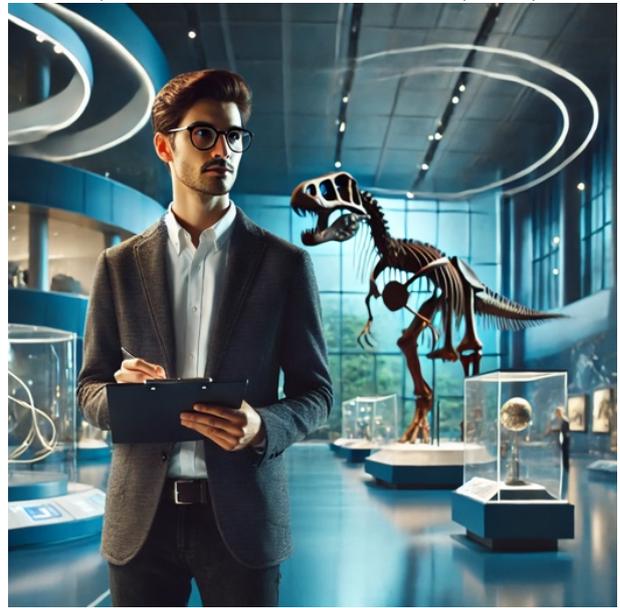

c) curator in a fashion museum (FA15)    d) curator in a science museum (SC7)

*Figure 1. Examples of museum curators depicted by ChatGPT4o*

A separate dataset, generating 200 images of women working in the creative and cultural industries included 50 women curators [34]. These generative AI representations show the women curators as purely Caucasian (100%), predominantly young (80%), wearing a blazer with primarily open blouses (72%) and glasses (56%). All museums are depicted as nineteenth century buildings, witch the exception of science museums.

*Table 4. Characteristics of museums depicted by ChatGPT4o*

|  | MA | NH | AM | FA | SH | SC | TE |
|---|---|---|---|---|---|---|---|
| **Building design** | | | | | | | |
| 19th century | 100.0 | 100.0 | 98.0 | 84.0 | 46.0 | — | — |
| 20th century | — | — | 2.0 | 16.0 | 54.0 | 70.0 | 100.0 |
| 21st century | — | — | — | — | — | 30.0 | — |
| **Visitors** | | | | | | | |
| in background | 10.0 | 50.0 | 42.0 | — | 42.0 | 90.0 | 90.0 |
| n | 10 | 50 | 50 | 50 | 50 | 10 | 10 |

## 4. Discussion

As there is no systematic overview of the demographic composition of curators employed in the world's museums and galleries [35], a comparison of the representations by DALL-E with the real world is difficult.

In the United Kingdom, 49% of museum staff who identified their gender were women [36]. In the USA women made up 58.6% of archivists, curators, & museum technicians in the 2022 census (2014: 63.1%; 2017: 63.9%; 2020: 60.4%) [37] The representation of women Australian gallery and museum curators has been steadily increasing over the past two decades, making up 72% according to the 2021 census (1996: 54.3%; 2001: 60.7%; 2006: 63.7%; 2011: 67.0%; 2016: 67.6%) [38-42]. Thus, the heavily biased gender representation of the visualisations of curators by generative AI does not reflect the reality on the ground but perpetuates the stereotype that 'responsible' and 'influential' positions are held by men, a pattern also observed in other generative AI visualisations [8, 22].

The age profile of archivists, curators, & museum technicians in the USA, when concatenated into the classes young (≤34), middle aged (35–54) and old (≥55), showed a broadly even distribution in 2014 (30.3% : 30.0% : 39.8%), becoming more skewed in 2022, with a decrease in middle aged and an increase in older curators (31.9% : 26.4% : 41.7%) [37]. The Australian data show that over 50% of curators were middle aged with the remainder roughly evenly split between young and older curators (2021—22.3% : 52.0% : 25.9%). Overall, Australian gallery and museum curators are becoming an increasingly ageing workforce (≥55: 1996:12.4%; 2011: 22.4%; 2021: 25.9%) [38, 39, 43]. As in the case of the generative AI representation of curators' gender, the visualisations of the age profile are heavily skewed towards younger staff, which does not reflect the reality in the institutions, nor does it reflect ChatGPT4o's own 'understanding' of the curatorial universe (see below).

Few statistics exist for the ethnicity of museum curators. In the USA, 82.6% of archivists, curators, & museum technicians were Caucasian in 2022 (down from 86.1% in 2014) [37]. In the United Kingdom, of all museum staff who stated their ethnicity, 64% were identified as white [36]. The ethnic composition of the curatorial population depicted by ChatGPT / DALL-E was one hundred percent white/Caucasian. This representation is significantly higher than the actual representation in the USA ($\chi^2$= 18.605, df=1, $p$<0.0001). This is not surprising if we posit that the primary source of training data is derived from English-speaking, primarily US sources and that the composition of the red team membership providing output quality control and moderation during the training phase is drawn from North American experts [44].

Not only are the vast majority of curatorial staff represented as young Caucasians, but all curators are also depicted as elegantly dressed, even when wearing casual wear, with coiffed, stylish hair, and well-trimmed beards. Not only are the curatorial characters as presented by ChatGPT / DALL-E a far cry from the popular stereotype of curators as nerdy, conservative in dress and stuck in time (see introduction), but in these visualisations they generally resemble yuppie professionals or people portrayed in fashion advertising (male 'eye-candy') rather than curators in real life.

Based on these observations of dissonance, the question was put to ChatGPT4o to "describe a typical curator in terms of gender, ethnicity, age and attire" to assess whether the response conformed to the stereotypes that were visualised (see supportive material, 'conversations' A–D) [28]. Congruent with its training to make ChatGPT4o respond in a balanced fashion [45, 46], the answers noted a preponderance of Caucasian curators noting increasing numbers of other ethnic backgrounds as well as the (possible) existence of non-binary individuals. Stated were a male dominance in art and natural history museums, while women were strongly represented in social history and fashion museums. When asked whether these observations would inform the creation of an image of a 'typical curator,' ChatGPT4o responded that it "*wouldn't rely solely on outdated stereotypes (e.g., an older white male in a tweed jacket), but rather create a plausible, modern representation that reflects the growing diversity in the field*" ('conversation' A) and that it would "*use the description as a general guide but avoid reinforcing stereotypes*" ('conversation' C).

Clearly, there is a dissonance between the 'understanding' of curators by ChatGPT and the resulting visualisation. Analysis of the image prompts generated by ChatGPT showed that biases are generated at two steps of the image generation process: the initial step that is autonomously generated by ChatGPT, which may specify age, ethnicity of gender, or may remain silent on this, and the interpretation of the injected prompt by DALL-E [47].

**5. Conclusions**

The study set out to examine how the popular generative AI model, ChatGPT4o, interprets and visualises curators, and to what extent these visualisations reflect the popular stereotype of curators as boring or at least nerdy, conservative in dress and stuck in time rummaging through collections in the basement. Based on 230 individually generative AI created visualisations of curators in seven different museum types, the study found a strong bias toward depicting men, with only a handful of women included. The AI-generated representations contrast sharply with real-world demographics. Women are extremely underrepresented, despite making up a significant portion of museum professionals in the English-speaking world. The women portrayed were all young and professionally dressed. Most male curators were also shown as young, particularly in fashion and natural history museums, while social history, maritime, and technology museums included more middle-aged individuals. Curatorial attire varied, with art and social history curators more often in informal clothing, while natural history curators were dressed more formally.

The AI-generated representations contrast sharply with real-world demographics. Women, for example, are extremely underrepresented, despite making up a significant portion of museum professionals in the English-speaking world. Likewise, the young professionals dominated age structure of curators among the generative AI visualisations, while in reality the majority of curators is in their late 40s and 50s.

Stereotypical attributes, however, were consistently depicted. Name tags appeared mostly in science, natural history, and maritime museums but were absent in art and fashion museums. Curators in art and fashion museums were frequently depicted with clipboards, while those in technology museums often held tablets. While glasses appeared inconsistently, beards were common in curators from art, natural history, and maritime museums but rare in fashion museums. Museum settings followed traditional representations, favouring historic nineteenth-century buildings.

The findings highlight biases introduced at multiple stages of AI-generated content, shaping an inaccurate portrayal of museum professionals if the images are taken uncritically at 'face value'. There is a risk that generative AI visualisations, due to their copyright free nature, are used as a generic stand-in illustration in media items, web pages of public presentations, generating false impressions.

**Funding**: This research received no external funding.

**Institutional Review Board Statement**: Not applicable

**Data Availability Statement:** The original data presented in the study are openly available via ¶¶

**Conflicts of Interest**: The author declares no conflict of interest.